# Graphene Unit Cell Imaging by Holographic Coherent Diffraction

Jean-Nicolas Longchamp, Tatiana Latychevskaia, Conrad Escher & Hans-Werner Fink

Physik Institut der Universität Zürich, Winterthurerstrasse 190, CH- 8057 Zürich, Switzerland

**We have imaged a freestanding graphene sheet of 210 nm in diameter with 2 Angstrom resolution by combining coherent diffraction and holography with low-energy electrons. The entire sheet is reconstructed from just a single diffraction pattern displaying the arrangement of 660.000 individual graphene unit cells at once. Given the fact that electrons with kinetic energies of the order of 100 eV do not damage biological molecules, it will now be a matter to develop methods for depositing individual proteins onto such graphene sheets.**

Despite the vast amount of structural information data on molecules made available by crystallography, a strong desire for imaging just one individual molecule is emerging. It would allow observing different molecular conformations that remain undiscovered as long as averaging is involved. For a meaningful contribution to structural biology, any such tool for single molecule imaging has to exhibit a resolution power of at least 2 Angstroms which has not been achieved so far. The strong inelastic scattering cross-section of X-rays and high-energy electrons as employed in conventional electron microscopes inhibits accumulating a signal-to-noise ratio record before the molecule is destroyed. While future X-ray Free Electron Lasers (XFELs) with drastically enhanced brightness and reduced pulse duration time might eventually attain the goal of single molecule imaging [1], the current and foreseeable state-of-the art in XFEL performance requires averaging over at least 1 million molecules to achieve atomic resolution [2]. With electrons of 200 keV kinetic energy, radiation damage is less severe and it has been possible to image an individual carbon nanotube with atomic resolution using coherent diffraction [3]. With low-energy electrons in the range of 100 eV molecules as fragile as DNA [4] remain unperturbed even after an exposure to a total dose of at least 5 orders of magnitude larger [5, 6] than the permissible dose in X-ray or high-energy electron imaging [7, 8]. However, the recent holographic imaging of a single ferritin molecule using such low-energy electrons exhibits a resolution barely below one nanometer, not sufficient to reveal any interesting structural detail of this protein [6]. Below we show that in using a highly transparent support and a combination of holography and coherent diffraction imaging [9] a resolution of 2 Å is achieved. For coherent diffraction imaging using low-energy electrons, a parallel wavefront is needed and is formed by a micro-machined electron lens placed in front of the electron point source [10, 11].

Since freestanding graphene is highly transparent for low-energy electrons [12, 13] it is possible to perform holography and coherent diffraction with one and the same sample. In the following, we show that a planar low-energy electron wave front provides a high-resolution coherent diffraction pattern of a circular 210 nm diameter graphene sample exhibiting diffraction spots corresponding to 2.13 Angstrom. Since the phase of the scattered wave is not known a priori, we also record a hologram of the very same sample and hence gain coarse information about the phase distribution. The combination of this initial phase distribution with the single high-resolution diffraction pattern [9] allows to iteratively reconstruct the entire 210 nm diameter graphene sheet displaying about 660.000 unit cells at once. A schematic of the experimental setup with its two modes of operation, coherent diffraction and holography, is displayed in Fig. 1.

Ultra-clean graphene was prepared by the platinum metal catalysis method [14], resulting in a freestanding graphene sheet placed over a 210 nm diameter bore ion milled through a platinum covered SiN membrane of 50 nm thickness. An electron hologram of such a freestanding graphene sheet is shown in Fig. 2 together with a coherent diffraction pattern of the very same sample. In both experimental schemes, the source of the coherent electron beam consists of a sharp W(111) tip that is driven by a 3-axis piezo-manipulator with nanometre precision. In the holographic mode (Fig. 1) the tip is brought as close as 380 nm to the sample and in the coherent diffraction mode, the distance between the electron source and the micro-lens is approximately 8 microns. In order to collimate the divergent electron beam originating from the electron source, an electrical potential difference of typically 50 V is applied between the two electrodes of the micro-lens. The sample is positioned into the parallel beam approximately 200 microns behind the lens with the help of a nanopositioner with three translational and one rotational axis. The resulting diffraction pattern is collected at a 68 mm distant electron detector of 75 mm in diameter (Fig. 1) and captured by an 8000×6000 pixels CCD chip. The outermost diffraction spots in the pattern displayed in Fig. 2 correspond to 2.13 Angstroms as determined by the de Broglie wavelength of the electrons and the angle under which they appear on the detector. Due to the fact that the initial phase distribution is directly available from the hologram and that the diffraction pattern is sufficiently fine sampled, freestanding graphene can be reconstructed with a resolution of 2 Angstroms.

This is done in the following manner. The experimental records, 16-bit TIFF images with 8000×6000 pixels in size, are first cropped to 6000×6000 pixels. The multi-channel plate grid image is numerically filtered out of the hologram by blocking the characteristic peaks in the Fourier domain. Thereafter, the hologram is normalized by division with the background image [15] and iteratively reconstructed [16, 17]. A twin-image suppressed reconstruction is already achieved after 6 iterations.

In order to build up a high dynamic range image, three diffraction patterns taken with different exposure times are combined appropriately. Based on the fact that the diffraction pattern of a real-valued object must be centrosymmetric, the experimental data are symmetrised for further enhancement of the signal-to-noise ratio [18]. Accordingly, the intensity of each pair of centrosymmetric pixels is set to their averaged intensity. Next, the diffraction pattern is transformed to spherical coordinates and multiplied with an apodization function to smoothen the transition to zero at the edges. In order to enhance the number of pixels per unit cell in the reconstruction, the diffraction pattern is zero-padded to 10.000x10.000 pixels. Accordingly, the hologram reconstruction is also re-sampled to 10.000x10.000 pixels. The overexposed region in the centre of the diffraction pattern is replaced by the corresponding 100 pixels in radius region of the squared amplitude of the Fourier transform of the hologram [9]. The iterative routine for the reconstruction is based on the Fienup algorithm [19]. The phase distribution used for the first iteration is provided by the phase of the Fourier transform of the hologram reconstruction [9]. The constraint applied in the object domain is positive absorption [16]. After the $50^{th}$ iteration, a mask is applied to set the transmission function outside the graphene covered bore to zero [17]. In the detector plane the constraint is just the basic physics concept that the amplitude must be equal to the square root of the measured intensity. After 100 iterations, the entire 210 nm diameter freestanding graphene sheet is recovered with 2 Angstrom resolution sufficient to display the roughly 660.000 graphene unit cells (Fig. 3 left). Since there is no way to print or display the entire reconstructed region of 10.000×10.000 pixels here, we have cut out and magnified some selected square areas of 5 nm side length [20]. The graphene unit cell is clearly apparent along with defects and domain boundaries present in the graphene sheet (Fig. 3 right). This is anticipated since the graphene sample is CVD grown on a polycrystalline copper substrate. In Fig. 4, we show another cut-out region with a size of 7.4×7.4 $nm^2$ where we have marked some apparent lattice defects in red.

To conclude, we have demonstrated that low-energy electrons allow damage-free imaging of a 210 nm diameter circular graphene sheet with 2 Angstrom resolution. A single diffraction pattern, if sampled at a sufficiently high rate, provides a real space image of more than half a million unit cells at once. Regarding freestanding graphene as support for molecules to be analysed, it will now be an immediate challenge to place a single protein onto such sample carrier and image it with the 2 Angstrom resolution.

**Acknowledgements:**

J.-N.L. performed the experiments and acquired the holograms and diffraction patterns. T.L. developed the numerical routines and performed the reconstruction of holograms and diffraction patterns. C.E. and H.-W.F. contributed to the design of the experimental setup and to sample preparation. All authors jointly drafted the manuscript.

The work presented here has been financially supported by the Swiss National Science Foundation (SNF).

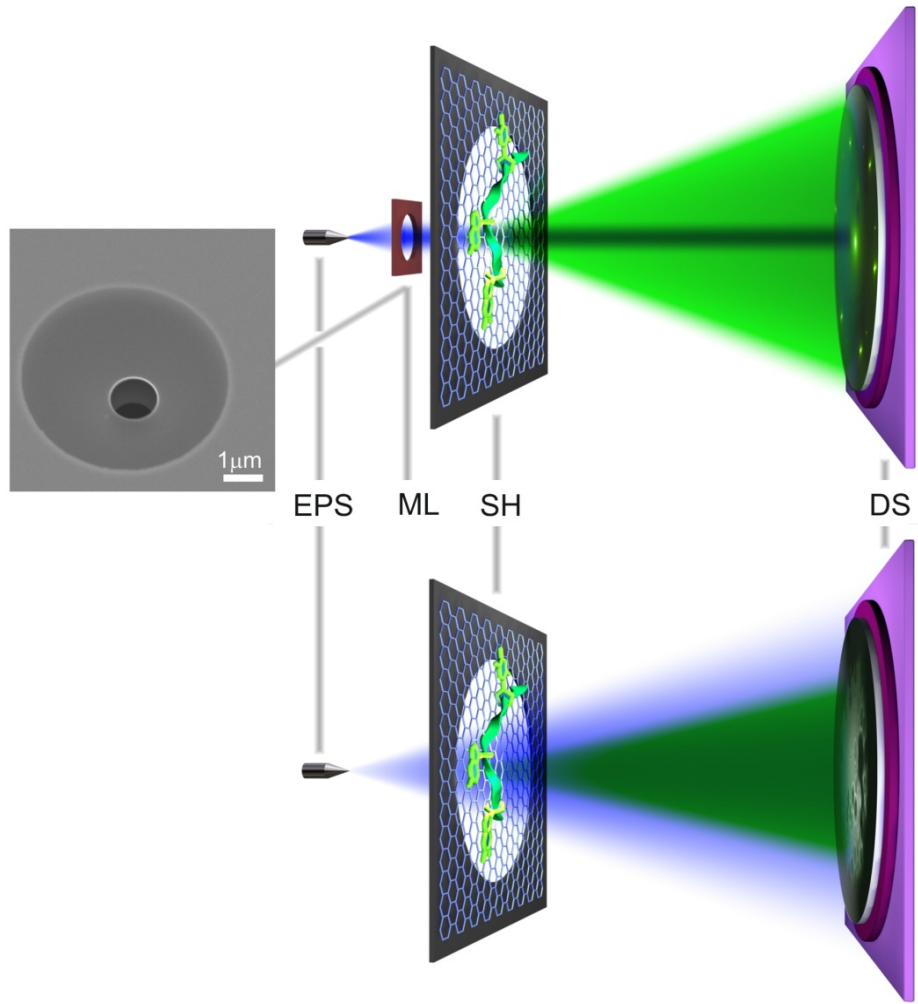

FIG. 1. Experimental setup for coherent diffraction and holographic imaging of freestanding graphene. An electron point-source (EPS) emits a spherical electron wave. For coherent diffraction imaging (top), a micro-lens (ML) with a bore of 1 micron (see inset showing a SEM image of the lens) is employed to form a parallel beam impinging onto the freestanding graphene mounted onto a sample holder (SH). The diffraction patterns, respectively the holograms (bottom), are recorded at a 68 mm distant detector system, consisting of a 75 mm diameter micro-channel-plate, followed by a phosphorous coated fibre-optic plate and an 8000x6000 pixels CCD chip.

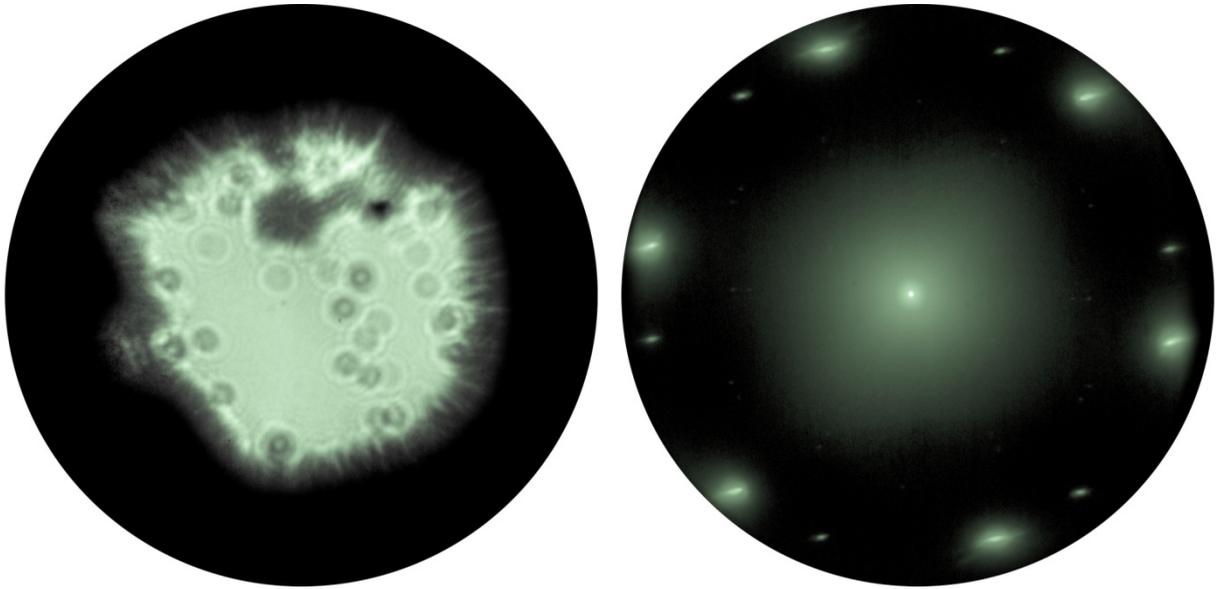

FIG. 2. Low-energy electron hologram and coherent diffraction pattern of freestanding graphene. Left: Hologram of freestanding graphene placed over a 210 nm diameter bore in a Pt covered SiN membrane, recorded with 58 eV electrons. Right: 236 eV electrons coherent diffraction pattern of the very same sample.

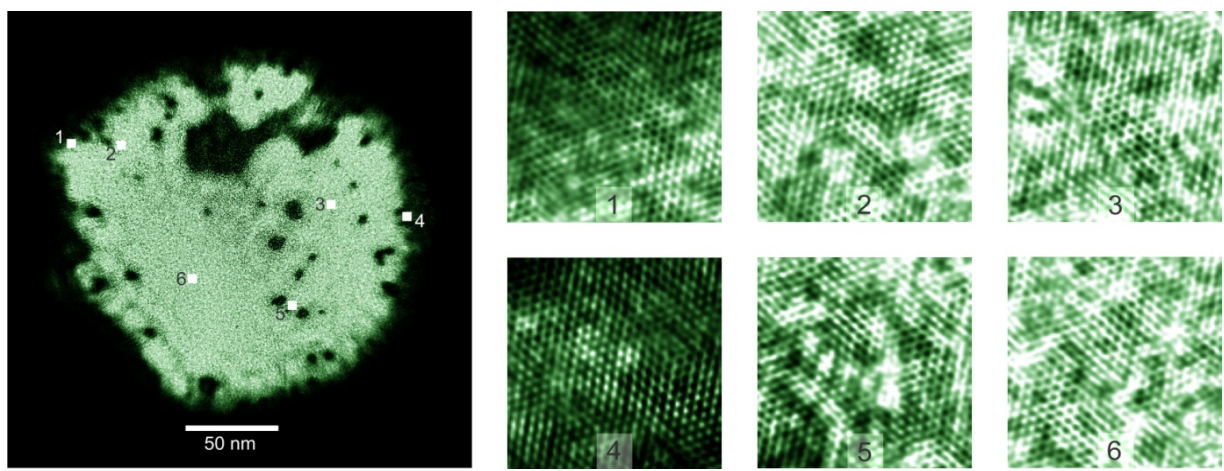

FIG. 3. Reconstruction of a 210 nm diameter freestanding graphene sheet with 2 Angstrom resolution. Left: reconstruction of the entire 210 nm diameter graphene sheet labelled with squares of 5x5 nm$^2$. Right: detailed view of the marked region in the left panel.

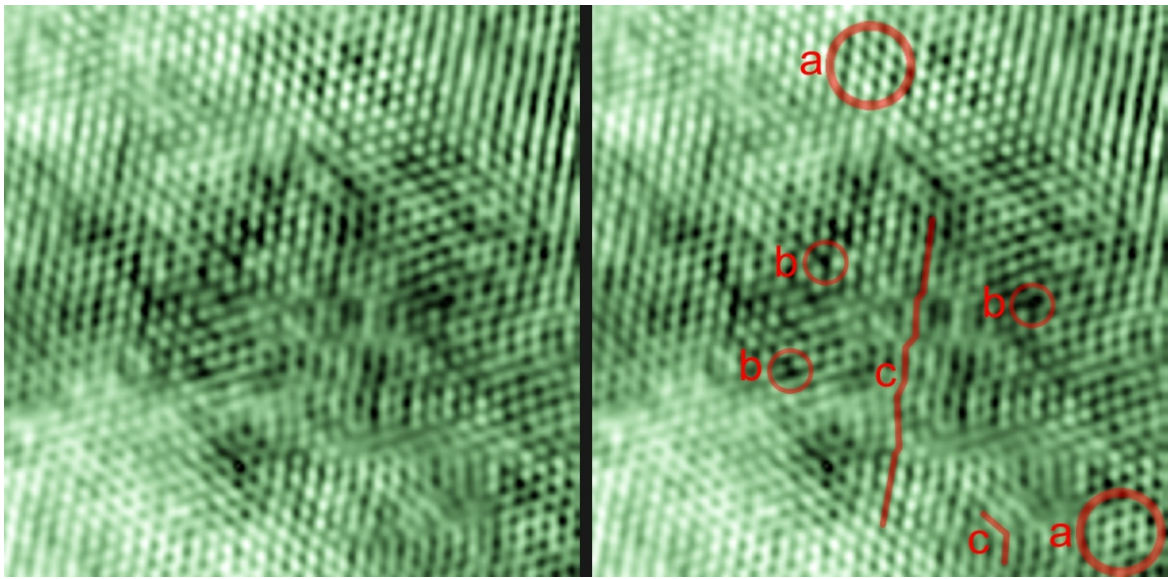

FIG. 4. Two identical copies of a 7.4×7.4 nm$^2$ region of the freestanding graphene sheet exhibiting various defects of the CVD grown graphene, marked in the copy at right with red circles respectively lines: (a) regions displaying perfectly arranged unit cells; (b) point defects; (c) domain boundaries.